\documentclass[pra, showpacs, twocolumn]{revtex4}

\usepackage{graphicx}
\usepackage{amsmath}

\begin{document}

\title{Tests for non-randomness in quantum jumps}

\author{D. J. Berkeland}
\email{djb@lanl.gov}
\author{D. A. Raymondson}
\altaffiliation[Present address:  ]{Physics Department, University
of Colorado, Campus Box 390, Boulder ,CO, 80309}
\author{V. M. Tassin}
\affiliation{Los Alamos National Laboratory, P-21, Physics
Division, MS D454, Los Alamos, NM 87545}

\date{\today}

\begin{abstract}
In a fundamental test of quantum mechanics, we have observed
228~000 quantum jumps of a single trapped and laser cooled
$^{88}$Sr$^+$ ion. This represents a statistical increase of two
orders of magnitude over previous similar analyses of quantum
jumps.  Compared to other searches for non-randomness in quantum
mechanical processes, using quantum jumps simplifies the
interpretation of data by eliminated multi-particle effects and
providing near-unit detection efficiency of transitions.  We
measure the fractional reduction in the entropy of information to
be $< 6.5~10^{-4}$ when the value of any interval between quantum
jumps is known. We also find that the number of runs of
successively increasing or decreasing interval times agrees with
the theoretically expected values. Furthermore, we analyze 238~000
quantum jumps from two simultaneously confined ions and find that
the number of apparently coincidental transitions is as expected.
Finally, we observe 8400 spontaneous decays of two simultaneously
trapped ions and find that the number of apparently coincidental
decays from the metastable state agrees with the expected value.
We find no evidence for short- or long-term correlations in the
intervals of the quantum jumps or in the decay of the quantum
states, in agreement with quantum theory.
\end{abstract}

\maketitle

An axiom of quantum mechanics is that we cannot predict the result
of any single measurement of an observable of a quantum mechanical
system in a superposition of eigenstates. Testing this principle
is important, not only for basic science, but also for
applications such as quantum random number generators (QRNG's) and
potential quantum computers. It is therefore surprising that in
spite of the many experiments sensitive to quantum mechanical
effects, only a few experiments have explicitly searched for
non-random behavior in long sequences of repeated quantum
measurements. In~\cite{Thomas:2000, Andre:2000}, the randomness of
the path of a single photon after a beamsplitter was used to build
QRNG's. In ~\cite{Silverman:1999}, the arrival times of decay
products of unstable nuclei were used to test the statistics of
quantum decay. Although both these systems rapidly give excellent
statistics, detector inefficiencies limit the conclusions that can
be drawn regarding the unpredictability of quantum mechanical
measurements. Furthermore, both systems are insensitive to certain
types of non-random behavior: averaging over many particles in a
collection of nuclei could obscure non-random behavior of single
systems, patterns in emission times or photon arrival times could
be overlooked because of inefficient detectors, and because
beamsplitters are always somewhat biased, QRNG's based on them are
designed to be insensitive to consecutive runs of transmissions or
reflections. All these problems can be avoided by observing the
times of quantum jumps in a single atom~\cite{Cook:1990} because
transitions between atomic levels can be detected with near-unit
efficiency with no multi-particle effects~\cite{Erber:1985}.
\begin{figure}
\includegraphics{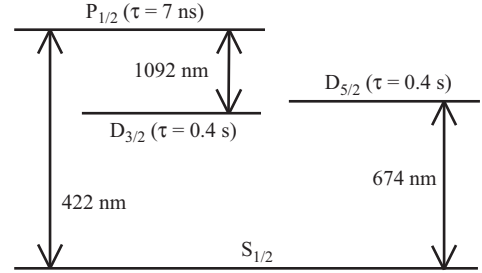}
\caption{\label{fig:1} Partial energy level diagram, transitions
and lifetimes in $^{88}$Sr$^{+}$.  We drive the 422-nm transition
to Doppler cool and detect the ions, and the 1092-nm transition to
prevent optical pumping into the D$_{3/2}$ state.  A laser of
bandwidth $<10$~kHz drives the 674-nm transition to produce
quantum jumps to and from D$_{5/2}$ states.}
\end{figure}

Here, we analyze the intervals between quantum jumps of a trapped
ion and answer the question ``after we have measured the length of
one interval, how much have we decreased the uncertainty of the
value of any interval we subsequently measure?" This question is
directly relevant to quantum logic operations on any quantum
mechanical system \cite{Galindo:2002}; if the result of one logic
operation affects that of subsequent (and supposedly independent)
operations then the operation of the gate is compromised. It is
also critical for demonstrating the suitability of a quantum
mechanical system as a random number generator.

Earlier work \cite{Erber:1989} examined 640 quantum jumps in a
single $^{198}$Hg$^+$ ion confined in a Paul
trap~\cite{Bergquist:1986}. Ref.~\cite{Erber:1995} later reported
a different but limited analysis of 10~000 quantum jumps. In a
substantial increase in statistics over this previous work, we
analyze 228~000 quantum jumps from a single ion comprising
continuous data sets of $\sim $10~000 events each.  We take
advantage of our greater statistics to also test for unexpected
correlations between transition times of multiple ions, analyzing
238~000 quantum jumps and 8400 spontaneous decays of pairs of
trapped ions.

\begin{figure}
\includegraphics{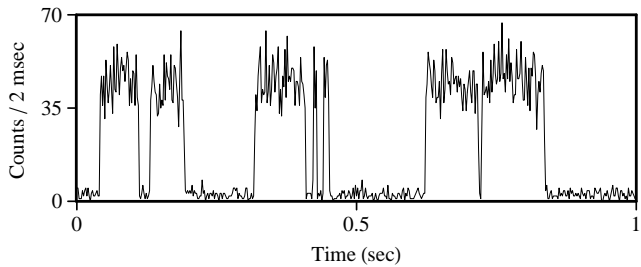}
\caption{\label{jumps} Number of 422-nm photons counted by the
detector over time, when all of the transitions shown in
Fig.~\ref{fig:1} are driven simultaneously.}
\end{figure}

We confine $^{88}$Sr$^+$ ions in a linear Paul
trap~\cite{Berkeland:2002} and simultaneously drive the
transitions shown in Fig.~\ref{fig:1}.  Figure~\ref{jumps} shows a
sample of the measured 422-nm scattering rate as a function of
time.  It displays the well-known characteristics of quantum
jumps: the ion rapidly scatters resonant 422-nm light when it is
in the S$_{1/2} \leftrightarrow$ P$_{1/2} \leftrightarrow$
D$_{3/2}$ manifold, but not at all when it is in one of the
metastable D$_{5/2}$ magnetic sub-levels. The scattering rate
changes abruptly whenever a 674-nm photon is absorbed or emitted.
According to quantum theory, the exact times of these changes, and
the intervals between them, should be unpredictable.

To test this assumption, we analyze the number of 422-nm photons
that scatter from the ion and reach a photomultiplier tube in a
measurement time $t_{meas}$ (typically a few ms, followed by a 200
$\mu$s dead time) when all three transition in Fig.~\ref{fig:1}
are simultaneously driven.  Each data set of approximately 10~000
measurements is taken over approximately 30 minutes.  From these
data we obtain four series $U = \{u_{1}, u_{2}, u_{3}, \ldots,
u_{N}\}$ of sequential time intervals, defined as follows. We
label as ``bright'' the series of intervals during which the ion
continuously scatters 422-nm light, and ``dark'' the series of
intervals during which the ion fluorescence is continuously
absent. In addition we analyze the series of intervals between
successive emissions (``emission'') and absorptions
(``absorption'') of a 674-nm photon.

Before each data set, we adjust the intensities of the laser beams
to control the rate of quantum jumps.  The bright intervals are
distributed exponentially, with average values ranging from 70 to
200~ms, while the average values for dark intervals range from 30
to 150~msec. When integrated over the long times of a data set,
the distributions of dark intervals deviate slightly from a purely
exponential form for interval durations greater than 100~msec.
This is due to drifts in the frequencies and intensities of laser
light at the site of the ion, which cause slight variations in the
average bright and dark interval durations. In addition, in the
presence of a degeneracy-breaking magnetic field $B$ (we vary $B$
from ~$10^{-6}$~T to $2~10^{-4}$~T), transitions from each
D$_{5/2}$ Zeeman sub-level contribute distinct exponential
distribution to the distribution of dark intervals. In spite of
their different responses to these two effects, the dark and
bright intervals in our experiment give statistically identical
results. We conclude from this that these effects neither
introduce nor mask correlations between intervals on the time
scales of our statistical tests.

Previous studies of randomness in quantum mechanics have used
several algorithms for answering the question regarding
predictability of intervals posed at this beginning of this paper:
for example, analyzing the distribution of sums and differences of
intervals~\cite{Erber:1989}, verifying that the probability of
particle decay is constant over short and long
time-scales~\cite{Silverman:1999}, and analyzing lengths of
continuously increasing and decreasing runs of
intervals~\cite{Silverman:1999, Erber:1989}.  We calculate the
difference between the entropy and conditional entropy of sets of
intervals, because this directly answers our question, and because
it results in an upper limit on the association of adjacent and
non-adjacent pairs of intervals.

To do this, we first divide our data sets into continuous strings
of 1000 events and normalize each interval time by the average
interval time of its string. This nearly eliminates effects due to
slow drifts of the rate of quantum jumps. Using each interval
$u_i$ only once, for $k \ge 1$ we tabulate the number of
occurrences $N_{m, n}$ of pairs \{$u_i$,~$u_{i+k}$\} for which
$u_i \in [t_m, t_{m+1})$ and $u_{i+k} \in [t_n, t_{n+1})$. We
choose $t_{m (n)}$ at intervals that result in roughly uniform
values of $N_{m, n} \gg 1$ for all $\{m, n\}$, although we find
that our results do not depend on the degree of uniformity. We set
the number of bins in each row and column ($m_{max} = n_{max}$) to
be 13.

Next, following Press~\cite{Press:1989} we call $p_{m \cdot}$ the
probability that the first interval in a pair ($u_i$) lies within
the range $[t_m, t_{m+1})$, and $p_{\cdot n}$ the probability that
the second interval ($u_{i+k}$) lies in the range $[t_n,
t_{n+1})$. We also call $p_{m, n} = N_{m, n}/N$ (where $N$ is the
total number of pairs) the probability that the pair falls in bin
$\{m, n\}$. The entropies are then defined as
\begin{equation}
\label{eq1} H(x) = -\sum\limits_n p_{\cdot n}~ln~p_{\cdot n}
\end{equation}
\begin{equation}
\label{eq2} H(y) = -\sum\limits_m p_{m \cdot}~ln~p_{m \cdot}~,
\end{equation}
\noindent and the conditional entropies are then defined as
\begin{equation}
\label{eq3} H(x|y)= -\sum\limits_{m,n} p_{m,n}~ln~(p_{m,n}/p_{m
\cdot})
\end{equation}
\begin{equation}
\label{eq4} H(y|x) = -\sum\limits_{m,n}
p_{m,n}~ln~(p_{m,n}/p_{\cdot n})~.
\end{equation}
\noindent These define two joint measures of the association
between the first and second intervals of any pair:
\begin{equation}
\label{eq5} U(x|y) = \frac{H(x) - H(x|y)}{H(x)}
\end{equation}
\begin{equation}
\label{eq6} U(y|x) = \frac{H(y) - H(y|x)}{H(y)}~.
\end{equation}

\noindent $U(x|y)$ represents the amount of information that the
distribution of first intervals gives about the distribution of
second intervals~\cite{Press:1989}, and vice versa for $U(y|x)$.
For an infinitely large data set, if the intervals in a pair are
statistically independent of each other then $U(x|y) = U(y|x) =
0$, and if they are completely associated with each other, $U(x|y)
= U(y|x) = 1$. However, statistical fluctuations in the
$N_{i,j}$'s limit how closely $U(x|y)$ and $U(y|x)$ approach 0.

\begin{figure}[b]
\includegraphics{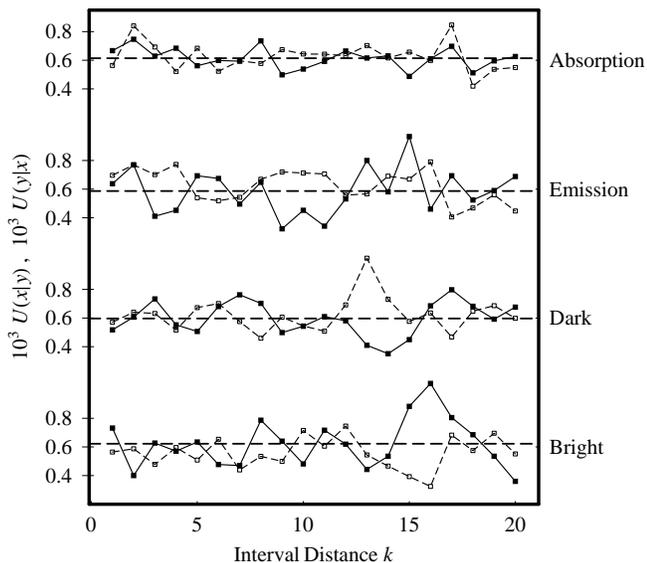}
\caption{\label{fig:4} The joint functions $U(x|y)$ (solid lines)
and $U(y|x)$ (dashed lines) from Eqs. \ref{eq5} and \ref{eq6} as a
function of the number of intervals between pairs for bright and
dark intervals and intervals between emissions and absorptions of
674-nm photons. Heavy dashed lines indicate the average values for
each interval type.}
\end{figure}

\begin{figure}
\includegraphics{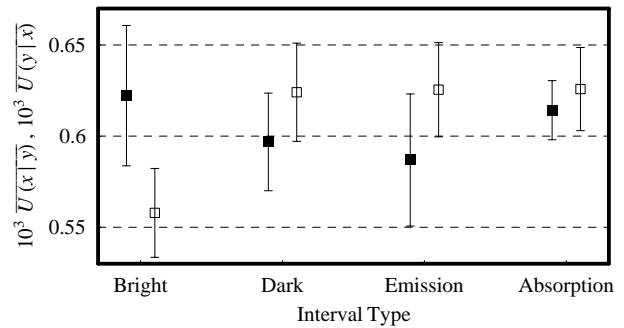}
\caption{\label{fig:6} The average values of $U(x|y)$ (solid
boxes) and $U(y|x)$ (hollow boxes) from Fig.~\ref{fig:4}.}
\end{figure}

Figure \ref{fig:4} shows $U(x|y)$ and $U(y|x)$ as functions of the
number of intervals $k$ that separate each pair $\{u_i, u_{i+k}\}$
for the four different interval types. The values do not depend on
$k$.  Figure~\ref{fig:6} shows the average values of $U(x|y)$ and
$U(y|x)$.  They are consistent with each other and with the value
0.00062 that is calculated if all of the $N_{m, n}$ were randomly
distributed about the average value $q = N~m^{-2}_{max}$ with a
distribution width $\sqrt{q}$. These facts show that intervals are
uncorrelated when they are separated by 1 to 20 intervals, which
corresponds to approximately 2~ms to several seconds.

\begin{figure}
\includegraphics{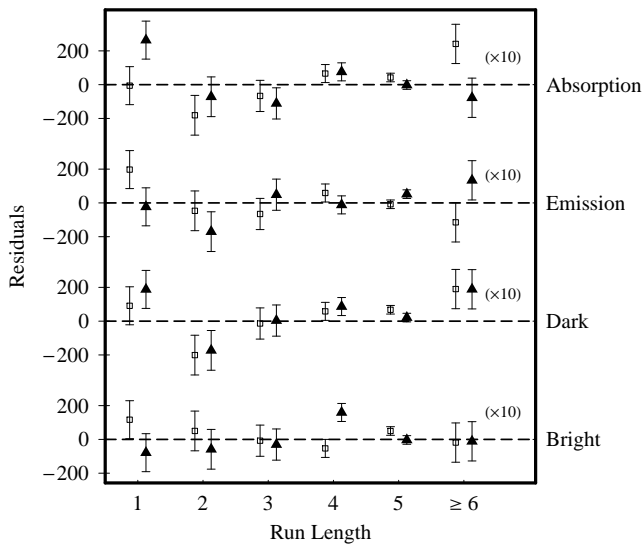}
\caption{\label{fig:5}  Differences between the expected and
measured values of the number of runs up (squares) and down
(triangles) for the four interval types.  The values of the number
of runs of length $\ell \ge 6$ have been magnified, and the runs
up and down have been slightly offset along the x-axis for
clarity. In total, each data set has $\sim$ 57~000 runs of various
lengths both up and down. }
\end{figure}

As a complimentary test for short- and long-term order, we search
for deviations from the expected distribution of consecutively
increasing or decreasing runs of elements of length $m$. As an
example, the subsequence $\{\ldots, 3, 15, 11, 5, 7, 2 \ldots \}$
contains a run down of three elements (15, 11 and 5) and a run up
of two elements (5 and 7). If the data set $U$ has a total of $N$
elements, the expected number of runs up or down of length $\ell$
is~\cite{Erber:1989}
\begin{equation}
\label{eq8} N_{runs} \left( {\ell} \right) = {\frac{{N\left(
{\ell^{2} + \ell - 1} \right) + \ell\left( {4 - \ell^{2}} \right)
+ 1}}{{\left( {\ell + 2} \right)!}}}~.
\end{equation}
\noindent This equation assumes that no element in $U$ is
repeated. However, we measure the intervals $u_i$ in integer
multiples of $t_{meas}$.  Often a run is terminated by a repeated
interval value, although it is equally likely that the run should
continue. We account for such cases by tallying two run lengths,
each with half the weight of a single run.  We make similar
adjustments when an interval occurs three times in a row. We
combine all of the data to obtain $N_{runs}(\ell)$ and their
deviation from the expected values, shown in Fig.~\ref{fig:5}.
Because the $N_{runs}(\ell)$ are not independent of each other, a
$\chi^2$ test does not apply. However, each point in the figure is
within reasonable agreement with the expected values.

A critical assumption in quantum algorithms involving multiple
quantum systems is that in the absence of intentionally applied
coupling, the individual systems do not communicate with each
other. We do not expect to observe cooperative effects in the
transitions times of pairs of atoms. However, other workers have
seen many more apparently simultaneous quantum jumps
\cite{Sauter:1986} and spontaneous decays~\cite{Block:1999} than
expected.  We take advantage of our statistics and search for
signs that transitions in ions stored simultaneously in the trap
are correlated.

We analyze 248~000 quantum jumps in two ions separated by $\geq
20~\mu m$, and count the number of times in which both ions appear
to change state during the same measurement time $t_{meas}$. In
such events, the number of detected 422-nm photons $n_{422}$ in
one measurement time is less than a threshold consistent with no
ions fluorescing, and is immediately followed or preceded by a
value of $n_{422}$ that is greater than a threshold consistent
with two ions fluorescing.  The probability of such events is
determined by the probability per unit time for a single ion to
change states, the average 422-nm photon scattering rate per ion,
the two threshold values and the measurement time $t_{meas}$. We
also account for the possibility of misinterpreting the scattering
rate from a single ion for that of two ions due to insufficient
resolution between the count rate distributions of one and two
ions. The total number of coincidental transitions into the
D$_{5/2}$ states is expected to be 308, and we measure 320. Also,
a total of 316 coincidental transitions out of the D$_{5/2}$
states is expected from our data, and we measure 313. In addition,
we find that the observed numbers agree with those produced in
Monte Carlo simulations of the data.

We also analyze the spontaneous decay of simultaneously trapped
ions, which may be more sensitive to effects such as interactions
with background gas. A brief ($\le$~0.2~s) saturating pulse of
674-nm light excites the atoms to the D$_{5/2}$ states while the
422-nm light is absent. After the 674-nm light pulse, the 422-nm
light is returned and we monitor the 422-nm photon scattering rate
every $t_{meas}$ = 5 msec. We observe 8400 decay processes that
start with two ions in the D$_{5/2}$ state and finish with no ions
in the D$_{5/2}$ state. In total, 26 of these transitions appear
to occur during the same measurement time $t_{meas}$. From a
measured decay rate of 410 msec in our system, we expect to see 24
(4) of these processes. This, too, is consistent with the behavior
of the ions being random, and agrees with the results
of~\cite{Steane:2000}

In conclusion, while it is impossible to \textit{prove}
randomness, we have seen no signs of non-random behavior over
short and long time scales after analyzing 228~000 quantum jumps
in single ions, 238~000 quantum jumps in two simultaneously
trapped ions, and 8400 spontaneous decays of two ions. Practical
QRNG's and quantum computers would use fewer quantum interactions
than those analyzed here. The present sensitivity is sufficient to
show that these applications are not affected by correlations due
to non-randomness of quantum mechanics.  However, at times short
compared to the transit time of light across the
ion~\cite{Enaki:1996}, and times long compared to the decay rate
of the atomic state ~\cite{Milonni:1994}, the Weisskopf-Wigner
approximation breaks down so that the distribution of decay times
is no longer exponential. In this case, we would expect to see
non-random behavior in the interval times of quantum jumps.
Observations in these regimes are out of the reach of modern
experiments, but are an intriguing possibility for future work.

This work was funded by DOE through the LDRD program. We would
like to thank Malcolm Boshier for carefully reading this
manuscript and for valuable discussions, and Richard Hughes for
initially bringing this topic to our attention.


\begin{thebibliography}{}
\bibitem{Thomas:2000} T. Jennewein, U. Achleitner, G. Weihs, H. Weinfurter and A. Zeilinger, Rev. Sci. Inst. \textbf{71}, 1675-1680 (2000).
\bibitem{Andre:2000}A. Stefanov, N. Gisin, O. Guinnard, L. Guinnard and H. Zbinden, J. Mod. Opt. \textbf{47}, 595-598 (2000).
\bibitem{Silverman:1999} (a) M. P. Silverman, W. Strange, C. Silverman, and T. C. Lipscombe,
Phys. Rev. A. \textbf{61}, 042106 (2000); (b) M.P. Silverman and
W. Strange, Phys. Lett. A \textbf{272}, 1-9 (2000).\par
\bibitem{Cook:1990} R. Cook, in \emph{Progress in Optics XXVII}, ed. E. Wolf (Elsevier Science Publishers,
B.V., 1990), pp. 362-416.
\bibitem{Erber:1985} T. Erber and S. Putterman, Nature \textbf{318}, 41-43 (1985).
\textit{Academic Press, Inc., Boston, 1994}, p. 148.
\bibitem{Erber:1989} T. Erber, P. Hammerling, G. Hockney, M. Porrati and S. Putterman,
Ann. Phys. \textbf{190}, 254-309 (1989).
\bibitem{Bergquist:1986} J.C. Bergquist, R.G. Hulet, W.M. Itano and D.J. Wineland,
Phys. Rev. Lett. \textbf{57}, 1699-1702 (1986).
\bibitem{Erber:1995} T. Erber, Ann. N.Y. Acc. Sci. \textbf{755}, 748-756 (1995).
\bibitem{Galindo:2002} A. Galindo and M.A. Mart\'{\i}n-Delgado, Rev. Mod. Phys. \textbf{74}, 347-423 (2002).
\bibitem{Berkeland:2002} D.J. Berkeland,  Rev. Sci. Inst. \textbf{73}, 2856-2860 (2002).
\bibitem{Wayne:1990} W.M. Itano, J.C. Bergquist, F. Dietrick, and D.J. Wineland,
in \emph{Coherence and Quantum Optics VI}, ed. J.H. Eberly
\textit{et al.} (Plenum Press, New York, 1990), pp. 539-543.
\bibitem{Press:1989} W.H. Press, B.P. Flannery, S.A.
Teukolsky and W.T. Vetterling, \emph{Numerical Recipies (FORTRAN
version)}, Cambridge Univerisity Press, Cambridge, 1989), p. 482.
\bibitem{Sauter:1986}Th. Sauter, R. Blatt, W. Neuhauser and P.E. Toschek,  Opt. Comm. \textbf{60}, 287-292 (1986).
\bibitem{Block:1999} M. Block, O. Rehm, P. Seibert and G. Werth, E.P.J.D. \textbf{7}, 461-465 (1999).
\bibitem{Steane:2000} C.J.S. Donald, D.M. Lucas, P.A. Barton, M.J. McDonnell, J.P. Stacey,
D.A. Stevens, D.N. Stacey and A.M. Steane, Europhys. Lett.
\textbf{51}, 388-394 (2000).
\bibitem{Enaki:1996} N.A. Enaki, JETP \textbf{82}, 607-615 (1996).
\bibitem{Milonni:1994} P.W. Milonni, \emph{The Quantum Vacuum},
(Acedemic Press, Inc., Boston, 1994), pp. 144-148.
\end{thebibliography}
\end{document}